\documentclass[fleqn,usenatbib,useAMS]{mnras}

\usepackage{mathptmx}

\usepackage[T1]{fontenc}

\usepackage{graphicx}	
\usepackage{amsmath}	
\usepackage{amssymb}	

\newcommand{\kev}{keV}

\newcommand{\nustar}{\textit{NuSTAR}}
\newcommand{\xmm}{\textit{XMM-Newton}}

\newcommand{\fe}{Fe~K$\alpha$}

\title[Sustaining a Warm Corona in AGNs]{Sustaining a Warm Corona in Active Galactic Nuclei Accretion Discs}

\author[D. R. Ballantyne \& X. Xiang]{
D. R. Ballantyne\thanks{E-mail: david.ballantyne@physics.gatech.edu}
and X. Xiang
\\
Center for Relativistic Astrophysics, School of Physics, Georgia
  Institute of Technology, 837 State Street, Atlanta, GA 30332-0430, USA\\
}

\date{Accepted XXX. Received YYY; in original form ZZZ}

\pubyear{2020}

\begin{document}
\label{firstpage}
\pagerange{\pageref{firstpage}--\pageref{lastpage}}
\maketitle

\begin{abstract}
Warm coronae, thick ($\tau_{\mathrm{T}}\approx 10$--$20$, where $\tau_{\mathrm{T}}$ is the
Thomson depth) Comptonizing regions with temperatures of $\sim
1$~\kev, are proposed to exist at the surfaces of
accretion discs in active galactic nuclei (AGNs). By combining with the
reflection spectrum, warm coronae may be responsible for producing the
smooth soft excess seen in AGN X-ray spectra. This paper
studies how a warm corona must adjust in order to sustain the
soft excess through large changes in the AGN flux. Spectra from one-dimensional constant density and
hydrostatic warm coronae models are calculated assuming the illuminating hard
X-ray power-law, gas density, Thomson depth and coronal heating
strength vary in response to changes in the accretion rate. We identify models that produce warm
coronae with temperatures between $0.3$ and $1.1$~\kev, and
measure the photon indices and emitted fluxes in the
$0.5$--$2$~\kev\ and $2$--$10$~\kev\ bands. Correlations and
anti-correlations between these quantities depend on the evolution and
structure of the warm corona. Tracing the path that an AGN follows
through these correlations will constrain how warm coronae
are heated and connected to the accretion
disc. Variations in the density structure and coronal heating strength
of warm coronae will lead to a variety of soft excess strengths and
shapes in AGNs. A larger accretion rate will, on average, lead to a
warm corona that produces a stronger soft excess, consistent with
observations of local Seyfert galaxies.

\end{abstract}




\section{Introduction}
\label{sect:intro}
One of the goals of studying the X-ray spectra of active galactic
nuclei (AGNs) is to understand the physics of accretion discs around
compact objects. Although the X-ray luminosity comprises only $\approx
10$--$20$\% of the bolometric luminosity produced by the disc \citep[e.g.,][]{elvis94,vf07,netzer19,duras20}, the X-rays
are produced close to the center of the disc ($\la 10$ gravitational
radii; e.g., \citealt{kara16}), in close proximity to where dissipation in the disc is expected
to peak \citep[e.g.,][]{ss73,fkr02}. The hard X-ray power-law is likely produced outside
the bulk of the disc in a hot, optically-thin corona \citep[e.g.,][]{galeev79,hm91,hm93,sz94,petrucci01}, and will
irradiate the disc surface, probing its metallicity, density
and dynamical structure \citep[e.g.,][]{rf93,brf02,fr10,garcia13,ball17}. Examining the details of the power-law
itself can also reveal how energy from the disc is deposited in the
hot corona \citep[e.g.,][]{fabian17}. Below $\approx 2$~keV most AGNs exhibit an excess of X-ray emission above
what is expected from extrapolating the power-law to lower energies \citep[e.g.,][]{wf93,gd04,bianchi09,gw20}. The origin of this `soft
excess' is still not understood, but this emission must be connected
to the energy released by the accretion disc. Therefore, the AGN soft
excess may provide another window through which to view the underlying
physics of accretion flows.

In recent years two alternative models for the soft excess in AGNs have been
debated in the literature. In one scenario, the excess is produced by
the sum of recombination lines and bremsstrahlung emission
from the heated surface of a disc that is illuminated by the external
power-law \citep[e.g.,][]{garcia19}. If the irradiated disc is close to the central black hole
(which it must be in many cases; e.g., \citealt{rm13}) then relativistic effects will
smear and broaden the emission so that it appears as a featureless
continuum arising out of the observed power-law \citep{crummy06}. This scenario
also naturally explains the soft X-ray reverberation lags observed in
some AGNs \citep[e.g.,][]{lobban18,mallick18,demarco19}. In the second
model, the soft excess is produced by Comptonization of thermal disc
photons by a warm ($\sim 1$~keV) and optically thick
($\tau_{\mathrm{T}} \approx 10$--$20$) layer of gas situated at the
surface of the disc \citep[e.g.,][]{mag98,czerny03,kd18,petrucci18}. This `warm corona' is distinct from the hot
corona that produces the hard X-ray power-law, but both must be heated
by energy released in and transported from the accretion disc. Broadband
X-ray spectra of bright AGNs can be successfully fit by both models
\citep[e.g.,][]{petrucci18,garcia19}, with the debate focusing on which scenario is the most
physically plausible \citep{garcia19}.

As the hard X-ray power-law appears to always be
present in AGNs \citep[e.g.,][]{liu17}, these two models are not
physically distinct and the irradiation from the power-law must be
included when calculating warm coronae properties. \citet{ball20}
modeled the physical conditions of warm coronae that included
the effects of the external power-law, and showed that a soft
excess can be produced by the \emph{combination} of both Comptonized
emission through a warm corona and reprocessed power-law radiation
(see also \citealt{kb16}). However, \citet{ball20} found that warm coronae with
temperatures necessary to generate a smooth, Compton dominated soft
excess exist in a `Goldilocks' zone of coronal heating and
cooling (see also \citealt{petrucci20}). The cooling rates of the gas are strong functions of density,
so if too much accretion energy is dissipated in the warm skin for a
given density, then the gas
overheats and the emitted spectrum is a Comptonized bremsstrahlung
spectrum extending to several keV. Alternatively, if the outer layers
of the disc are not heated enough then the gas can efficiently cool
and significant line emission is
predicted in the soft excess below 2~keV. Both of the overheating and
overcooling spectral shapes are inconsistent with observations, so if
a warm corona is an important contributor to the soft excess, tight constraints can, in principle, be placed on
the heat transport and dissipation processes in accretion discs.

Observational campaigns on bright AGNs show that the soft excess
remains a prominent feature in the spectrum over a wide range of flux
levels \citep[e.g.,][]{scott12,winter12,ricci17,gw20}. In addition, there is evidence that the shape and
strength of the soft excess is correlated with changes in the hard
X-ray power-law and the overall accretion rate through the
disc \citep[e.g.,][]{middei19,gw20}. Given that producing a warm corona appears to require
specific limits on the heating and cooling processes in order to
produce the necessary spectral shape, the observed persistence and
changes of the soft excess in AGNs provide additional constraints on
the applicability of warm coronae. In this paper, we extend the warm corona models of
\citet{ball20} to examine the requirements necessary to maintain a soft excess in
AGN spectra. We investigate changes in the
warm corona optical depth, density, irradiation conditions and heating
rate, and identify how these properties must vary in order to
maintain a soft excess in the observed spectra. Moreover, we measure how the
soft excess strength and shape varies as the warm corona parameters
change, which then can be tested against observations. As anticipated,
soft excesses can only be maintained by a warm corona if the parameters
describing the corona change in specific ways. The next section
outlines the calculations described in the paper, with results
presented in Sect.~\ref{sect:results} and discussed more broadly in
Sect.~\ref{sect:discuss}. The conclusions are presented in
Sect.~\ref{sect:concl}. 

\section{Calculations}
\label{sect:calc}
The warm corona models follow the procedure laid out by \citet{ball20} and
consists of a one-dimensional calculation of the thermal and ionization
balance of a constant density slab (with Thomson depth
$\tau_{\mathrm{T}}$) irradiated from above by a hard X-ray power-law
and from below by a blackbody. Following the warm corona hypothesis,
 a fraction $h_f$ of the energy dissipated by accretion ($D(R)$, where
 $R$ is the disc radius; e.g., \citealt{ss73}) is
 deposited into the layer with the following heating function (in erg cm$^{3}$
s$^{-1}$)
\begin{equation}
  \mathcal{H} = {h_f D(R) \sigma_{\mathrm{T}} \over \tau_{\mathrm{T}}
    n_{\mathrm{H}}},
  \label{eq:hfunct}
\end{equation}
where $\sigma_{\mathrm{T}}$ is the Thomson cross-section, and
$n_{\mathrm{H}}$ is the density of the slab. The X-ray reflection code of \citet{brf02} is used to solve the
thermal and ionization structure of the slab, and to predict the
X-ray spectra emitted from the surface of the gas, including the
effects of Comptonization \citep{ross79,rf93}.

The energy dissipated by accretion, $D(R)$, provides the total energy
budget available for irradiating and heating the slab. A fraction
$f_{\mathrm{X}}$ of the energy is transported outside the disc and released in
the hot corona. Therefore, the hard X-ray power-law impinging the
surface of the slab has a flux $f_{\mathrm{X}}D(R)$. The power-law has a
photon-index of $\Gamma=1.9$ and exponential cutoff energies at
$30$~eV and $220$~keV \citep{ricci17,ricci18}. As seen in Eq.~\ref{eq:hfunct}, a separate fraction
($h_fD(R)$) is released as heat in the slab, with a constant heating
rate per particle. Any remaining energy, $(1-h_f-f_{\mathrm{X}})D(R)$, is released
into the lowest zone of the layer as a blackbody with a temperature
given by the standard blackbody equation. This thermal emission is not
considered to be the source of the seed photons for the hard X-ray
power-law; therefore, the shape of the power-law is not affected by
the value of $h_f$. The physical picture is that
the computational domain corresponds to the top $\tau_{\mathrm{T}}$ at
the surface of an accretion disc at radius $R$, with the hot, X-ray
emitting corona physically distinct from this warm layer. We defer a
two-dimensional model with closer coupling between the hot and warm
coronae to future work. 

\citet{ball20} considered a fixed $D(R)$ and $f_{\mathrm{X}}$, and examined the
emitted spectra and physical conditions in the slab for different
$\tau_{\mathrm{T}}$ and $n_{\mathrm{H}}$. As mentioned above, a Compton
dominated warm corona with $kT \sim 1$~keV that leads to a reasonable
soft excess in the emitted X-ray spectrum can be produced as long as
the gas is neither too hot nor too cold. Therefore, if the `correct'
warm corona conditions are satisfied for a particular $D(R)$, corresponding to a
certain accretion rate through the accretion disc, then a large enough
change in $D(R)$ could substantially alter the heating and cooling
balance and lead to a layer too hot or cold to maintain a soft
excess. However, if one of the other parameters, such as $f_{\mathrm{X}}$,
$n_{\mathrm{H}}$ or $\tau_{\mathrm{T}}$, also varies with $D(R)$ in
just the right way, then they could compensate for the change in
$D(R)$ and maintain the soft excess in the emitted spectrum. Thus, if
maintaining a warm corona is necessary to explain AGN X-ray spectra,
this could lead to specific constraints on the behaviour of the
disc properties due to changes in the underlying accretion rate.

\section{Results}
\label{sect:results}
In this section, results are presented for experiments where the
energy distribution and properties of the irradiated slab are varied
in response to a change in the dissipated accretion power, $D(R)$. The
goal is to search for one or more relationships that will maintain a
$\sim 1$~\kev\ warm corona, and its resulting soft excess, as the flux
passing through the layer changes.

\subsection{Varying the Hot Corona Fraction $f_{\mathrm{X}}$}
\label{sub:fx}
A fraction $f_{\mathrm{X}}$ of the accretion power in the inner
accretion disc must be transported outside the disc and dissipated
in an optically thin, hot corona that produces the hard X-ray power-law
\citep[e.g.,][]{hm91,petrucci01}. \citet{ball20} found that the
irradiation from this power-law was important in the development of a
warm corona, as it provided a base level of heat and ionization in the
outer few Thomson depths of the heated slab. This background
ionization ensured that the soft excess emitted by the warm corona
was not swamped by absorption lines \citep[e.g.,][]{garcia19}.

Observations of AGNs have shown that the fraction of bolometric
luminosity that is emitted at energies $> 2$~\kev\ falls at larger
accretion rates \citep[e.g.,][]{duras20}. Thus, if a warm corona exists at some initial
accretion rate, corresponding to an initial dissipation $D_i(R)$ and
$f_{\mathrm{X}}$, then a change in accretion rate will lead to a change
in the external X-ray irradiation striking the disc. If $f_{\mathrm{X}}$
falls too far as $D(R)$ increases, then its possible it may no longer
provide enough ionization to ensure the absence of absorption lines;
similarly, if $f_{\mathrm{X}}$ rises too fast as $D(R)$ decreases,
then the external X-rays may provide too much heating and destroy the
warm corona.

\begin{figure*}
  \includegraphics[width=0.98\textwidth]{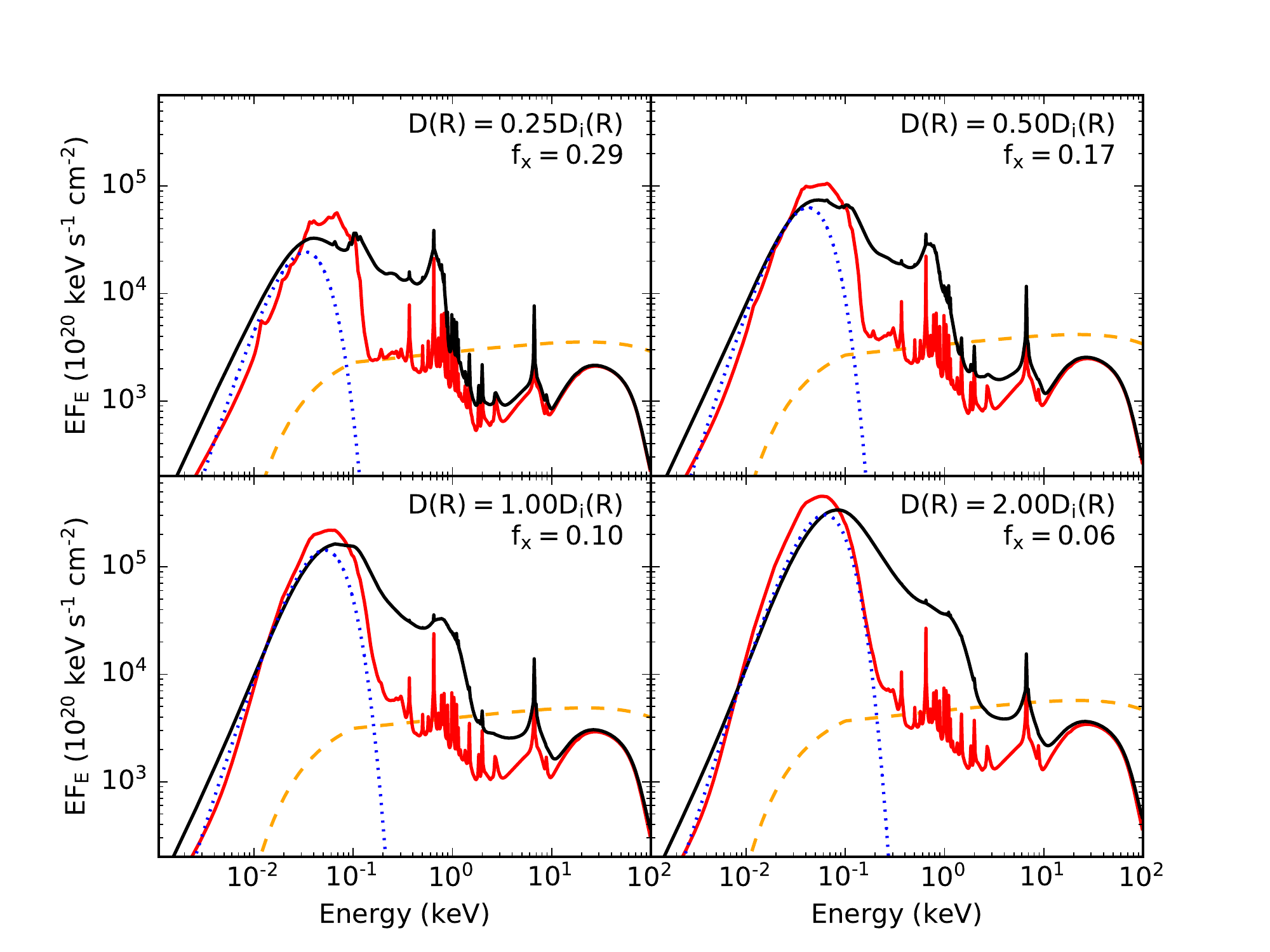}
  \caption{The black lines plot the spectrum emitted from the surface
    of a constant density slab with $n_{\mathrm{H}}=10^{14}$~cm$^{-3}$
    and $\tau_{\mathrm{T}}=10$ illuminated by a cutoff power-law from above
    (dashed orange line) and a blackbody (dotted blue line) from
    below. A total energy flux $D(R)$ is available to the model, with
    $f_{\mathrm{X}}D(R)$ going into the external power-law, $h_fD(R)$
    uniformly injected as heat into the slab (see
    Eq.~\ref{eq:hfunct}), and the blackbody taking up the
    remainder. Each panel in the figure shows the results for
    $h_f=0.3$ when $D(R)$ is varied above and below its initial value
    ($D_i(R)=5.175\times 10^{16}$~erg~cm$^{-2}$~s$^{-1}$) under the
    assumption $f_{\mathrm{X}} \propto D(R)^{-0.77}$ \citep{sr84}. The solid
    red line in each panel is the spectrum that results with no
    coronal heating. These plots show that $f_{\mathrm{X}}$ plays a
    minor role in the ability to maintain a soft excess in an AGN
    spectrum as $D(R)$ varies.} \label{fig:fxtest}
\end{figure*}

To check the resilience of the warm corona to changes in
$f_{\mathrm{X}}$, we calculated models of a heated slab with a fixed density
of $10^{14}$~cm$^{-3}$ and Thomson depth $\tau_{\mathrm{T}}=10$. The
initial accretion flux is $D_i(R)=5.175\times
10^{16}$~erg~cm$^{-2}$~s$^{-1}$, corresponding to the dissipated flux
(per disc side) at a radius of $5$ Schwarzschild radii from a
$6.25\times 10^6$~M$_{\odot}$ black hole accreting at $0.15\times$ its
Eddington rate \citep{sz94}. The hot corona fraction is set to
$f_{\mathrm{X}}=0.1$ at $D_i(R)$, and is assumed to vary as
$f_{\mathrm{X}} \propto D(R)^{-0.77}$ \citep{sr84}. With this setup as the
starting point, we calculate the emitted spectra with different coronal heating
fractions, $h_f$, for $D(R)/D_i(R)=0.25,
0.5, 1$ and $2$. The resulting spectra for $h_f=0.3$ are shown as the
solid black lines in Figure~\ref{fig:fxtest}.

The figure shows the impact of increasing and decreasing
$f_{\mathrm{X}}$ as a result of varying $D(R)$. The top-left panel
plots the case where nearly 30\% of $D(R)$ is released in the hot
corona. Despite the strong external illumination, which corresponds to
an ionization parameter ($\xi=4\pi f_{\mathrm{X}}D(R)/n_{\mathrm{H}}$)
of $1886$~erg~cm~s$^{-1}$, the surface
temperature of the slab is $\approx 0.35$~keV, and Compton cooling
does not dominate at any point in the slab. Thus, only a weak soft
excess, covered with recombination lines, is produced by this
model. As $D(R)$ increases, the flux illuminating the outer surface of
the slab falls, but the heating function $\mathcal{H}$
(Eq.~\ref{eq:hfunct}) grows, releasing more and more energy into the
slab. As a result, both the surface temperature and the importance of
Compton cooling in the gas increases with $D(R)$, generating a
stronger and smoother soft excess in the emitted spectrum. Indeed,
when $D(R)/D_i(R)=2$, Compton cooling dominates the entire slab and
the surface temperature reaches $\approx 1.3$~keV, exceeding the
values typically inferred from observations
\citep{petrucci18}. Clearly, when compared to changes in
$\mathcal{H}$, the value of $f_{\mathrm{X}}$ is not important in
  determining the strength or shape of the soft excess produced by a
  warm corona. However, the presence of the hard power-law is still
  required to maintain the base level of surface ionization and avoid
  strong absorption lines blanketing the emitted spectrum. In the
  following calculations, we continue to assume $f_{\mathrm{X}}
  \propto D(R)^{-0.77}$. 

\subsection{Varying the Density and Thomson Depth}
\label{sub:nhandtau}
The above experiment indicates that maintaining an appropriate warm
corona through changes in an AGN accretion rate must involve the
regulation of the coronal heating function, $\mathcal{H}$
(Eq.~\ref{eq:hfunct}). As found by \citet{ball20}, if $\mathcal{H}$ is
too large or too small, then the warm corona is either too hot or too
cold to produce a soft excess consistent with observations. Therefore, other quantities affecting $\mathcal{H}$ must
 adjust while $D(R)$ changes in order for a warm corona to continuously produce an AGN soft excess.

 Models of accretion discs provide some guidance on how discs adjust
 due to changes in accretion rates. \citet{jiang19} ran two global radiation
 magnetohydrodynamics simulations of AGN discs separated by nearly a
 factor of 3 in accretion rate. They find that the surface density in
 the inner part of the disc rises with the accretion rate. As a result,
 the disc density falls more slowly with height at the larger
 accretion rate. These results indicate that both the density and optical
 depth of a heated warm corona might increase in a disc with a larger
 accretion rate. In addition, \citet{jiang19} find that the dissipation fraction
 in optically thin material is reduced in the higher accretion rate
 simulation. This result is consistent with the drop in
 $f_{\mathrm{X}}$ inferred from observations, and could indicate that
 $h_f$, the fraction of accretion powered released in a warm corona,
 will increase as the accretion rate grows.

 To evaluate how the structural changes in the disc found by the
 simulations will impact maintaining an appropriate warm corona, we calculate
  the spectra produced by irradiated, heated slabs where
  $\tau_{\mathrm{T}} \propto D(R)^{3/5}$ and $n_{\mathrm{H}} \propto
  D(R)^{2/5}$. These relationships are taken from the predictions of
  gas-pressure dominated $\alpha$-discs \citep{sz94} and both
  qualitatively follow the
  same trends with accretion rate suggested by the numerical
  simulations of \citet{jiang19}. The baseline model has $D(R)=D_i(R)$,
  $n_{\mathrm{H}}=10^{14}$~cm$^{-3}$, $f_{\mathrm{X}}=0.1$, and
  $\tau_{\mathrm{T}}=8$. This value of $\tau_{\mathrm{T}}$ is
  consistent with the smallest optical depths inferred for warm
  coronae \citep{petrucci18}. From this baseline model we compute the
  properties of the slab for $D(R)/D_i(R)=1,2,4$ and $8$, with the
  density, optical depth, and $f_{\mathrm{X}}$ appropriately scaled
  using the above relationships. In addition, for each value of
  $D(R)$, the fraction of energy dissipated as heat into the slab,
  $h_f$, is varied from $0.1$ to $0.8$ (in steps of $0.1$) in order to
  consider a wide range of warm corona properties in each setup.

\begin{figure*}
  \includegraphics[width=0.98\textwidth]{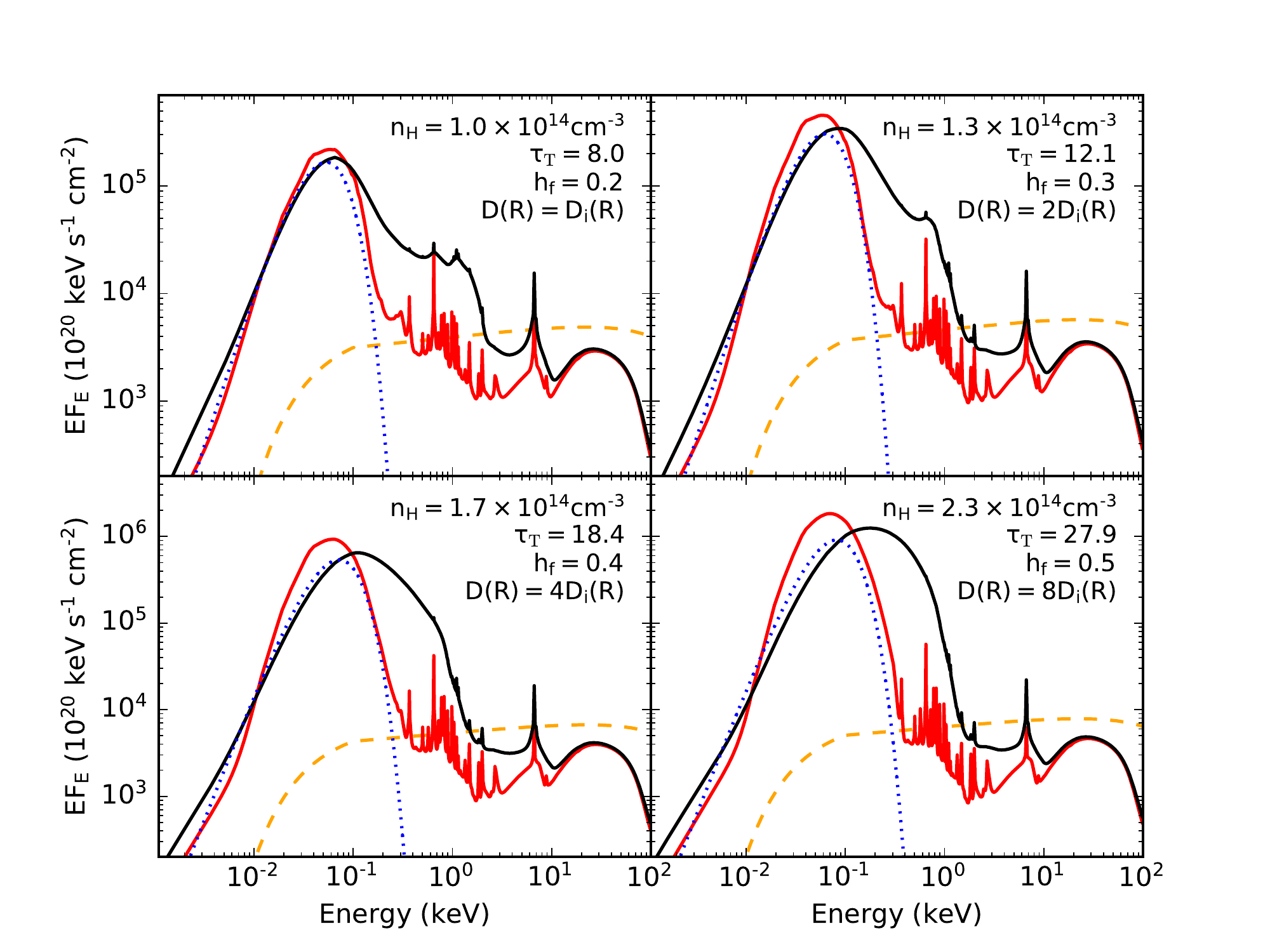}
  \caption{Predictions of the emitted spectrum from an illuminated constant density
    slab with density $n_{\mathrm{H}}$ and Thomson depth
    $\tau_{\mathrm{T}}$. In these models, the density, Thomson depth, and
    fraction of energy released in the hard X-ray power-law
    ($f_{\mathrm{X}}$) are all functions of the dissipated accretion
    energy $D(R)$ (see text). The plots show a scenario where $h_f$, the fraction
    of the accretion energy heating the warm corona, increases
    with $D(R)$ and maintains a Compton dominated soft
    excess. The line styles are the same as in Figure~\ref{fig:fxtest}.} \label{fig:constdens}
\end{figure*}

Figure~\ref{fig:constdens} shows the results from four models, one at
each value of $D(R)$, that produce a Compton-dominated warm corona and
a smooth soft excess.
The figures show a situation where as $D(R)$ increases, driving growth in both
$n_{\mathrm{H}}$ and $\tau_{\mathrm{T}}$, the heating fraction $h_f$
also grows to maintain a warm corona with similar
properties (see, e.g., Eq.~\ref{eq:hfunct}). The surface temperatures of the four slabs are not
widely different, ranging from $\approx 0.78$~keV for the $D(R)/D_i(R)=1$ model (top-left
panel) to $\approx 0.9$~keV for the remaining scenarios. The shape of
the soft excess predicted in these four models changes significantly
as the density and optical depth increases. In the $\tau_{\mathrm{T}}=8$
model, the soft excess still exhibits features due to spectral lines
and edges, and shows a modest Compton-scattered continuum formed from
the high-energy tail of the blackbody. However, as both
$\tau_{\mathrm{T}}$ and $n_{\mathrm{H}}$ increase, the spectral
features are erased from the soft excess, and a larger fraction of the
blackbody is Compton scattered to form a strong soft excess. This is a
result of the blackbody having to pass through a thicker
ionized layer as $D(R)$ increases. In fact, the critical optical
depth at which Comptonization saturates for a $10$~eV photon passing through a
$1$~\kev\ gas is $\tau_{\mathrm{T,crit}} \approx 28$, which
is nearly identical to the optical depth used in our models with
$D(R)/D_{i}(R)=8$.  Thus, we expect that models with larger values of
$\tau_{\mathrm{T}}$ will yield spectral shapes similar to the
$\tau_{\mathrm{T}}=27.9$ model shown here.

\begin{figure*}
  \includegraphics[width=0.98\textwidth]{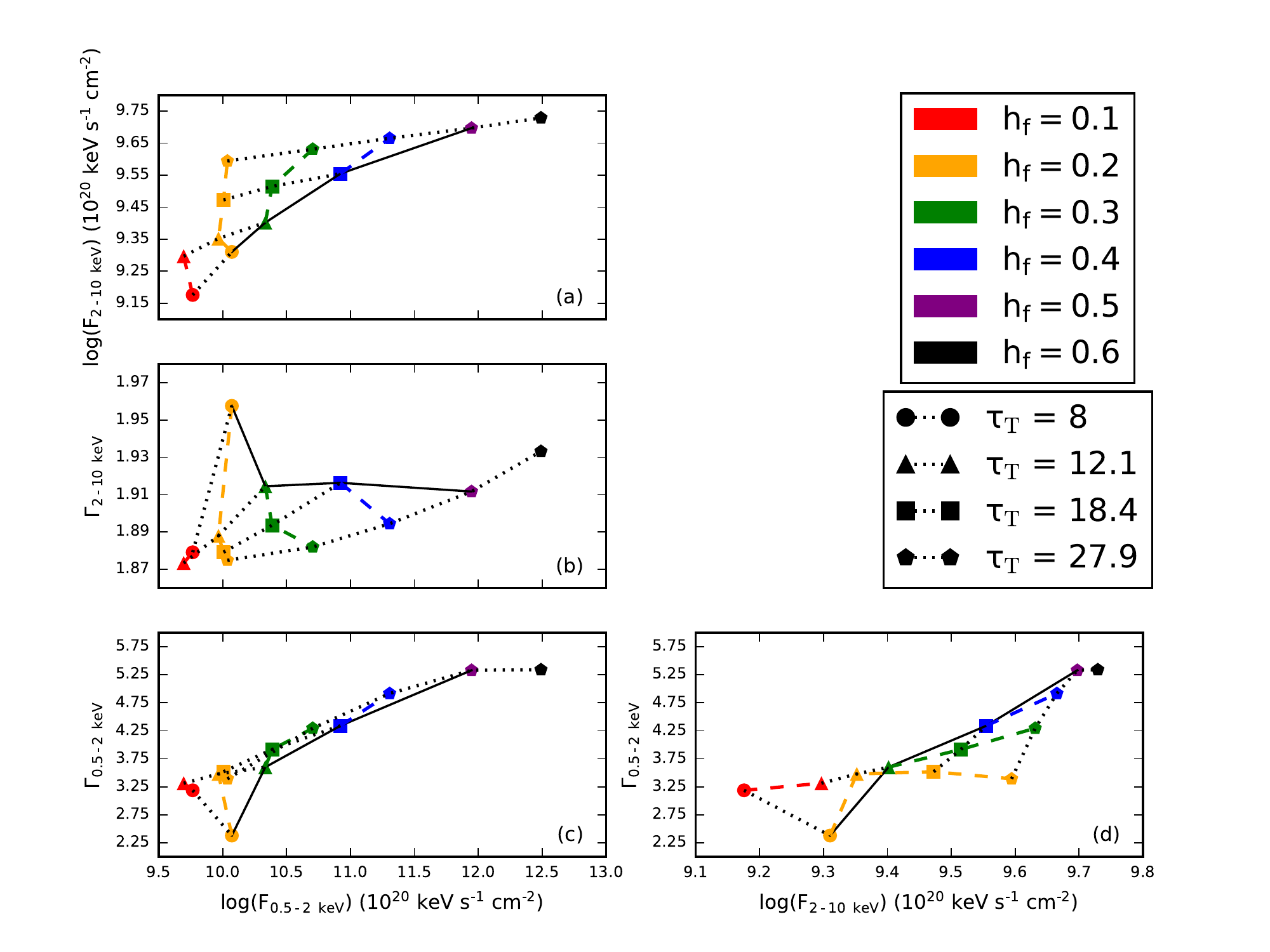}
  \caption{Trends of observational parameters in the X-ray spectra
    produced by the constant density warm corona models where
    $\tau_{\mathrm{T}}$, $n_{\mathrm{H}}$ and $f_{\mathrm{X}}$ are all
    functions of $D(R)$ (see text). Only models with a
    surface temperature between $0.3$ and $1.1$~\kev\ are included in
    this analysis to ensure they produce warm coronae consistent with
    observations \citep{petrucci18}. The photon indices and fluxes are
    calculated from spectra that combine the predicted
    emission and reflection spectrum (black solid lines in
    Fig.~\ref{fig:constdens}) with the illuminating power-law (dashed
    lines in Fig.~\ref{fig:constdens}), assuming a reflection fraction
    of $0.7$ \citep[e.g.,][]{zappa18}. The different symbols denote
    the results from the different $\tau_{\mathrm{T}}$ while the
    colours indicate the value of $h_f$ used in the model. Dotted lines
    in each panel connect results with the same $\tau_{\mathrm{T}}$,
    and the coloured dashed lines connect results with the same
    $h_f$. The solid black line in each panel traces out the path of
    the four models from Fig.~\ref{fig:constdens}. The fluxes are
    calculated in units of emitting area.  } \label{fig:scatterconstdens}
\end{figure*}

To determine how X-ray observations of AGNs may probe the effects of 
a changing $n_{\mathrm{H}}$, $\tau_{\mathrm{T}}$ and $h_f$ within warm
coronae, we add the illuminating power-law to the
predicted emitted and reflected spectrum in each model (assuming a reflection
fraction of $0.7$\footnote{i.e., Total Spectrum=Power-Law+$0.7\times$Reflection}; \citealt{zappa18}), and calculate the photon index
($\Gamma$) and flux in both the $0.5$--$2$~\kev\ and
$2$--$10$~\kev\ bands. The results for models with surface
temperatures in the range $0.3$--$1.1$~keV are plotted in
Figure~\ref{fig:scatterconstdens}. These temperatures select warm
corona models that span the range found by spectral modeling
\citep[e.g.,][]{petrucci18}. 
In order to isolate the effects of the changing parameters, different
symbols are used to indicate warm coronae with different
$\tau_{\mathrm{T}}$ and colours are used to distinguish models with
different values of $h_f$. Dotted lines connect models with equal
$\tau_{\mathrm{T}}$ and coloured dashed lines connect models with
equal $h_f$. An AGN with a soft excess produced by a constant density warm corona will be observed to move through the panels of
Fig.~\ref{fig:scatterconstdens} on a particular trajectory as its
accretion rate changes. For example,
the four models shown in Fig.~\ref{fig:constdens} trace out the solid
black lines in each panel.

Panel (a) of Fig.~\ref{fig:scatterconstdens} plots the predicted
$2$--$10$~\kev\ flux versus the $0.5$--$2$~\kev\ flux from the sample
of warm coronae. The fluxes are calculated directly from the total spectra and
are in units of emitting area. The distribution of points show that
warm coronae with a constant $\tau_{\mathrm{T}}$ (dotted lines) are
nearly orthogonal in this plane to those with fixed $h_f$ (coloured
dashed lines). Horizontal movement across the plot is largely driven
by increasing $h_f$ (i.e., releasing more energy in the warm corona) which
enhances the soft flux.  In contrast, increases in the hard X-ray
flux requires more total energy dissipated in the X-ray power-law,
which is most easily achieved by a larger $D(R)$ which, in turn, produces a larger $\tau_{\mathrm{T}}$. However, as $f_{\mathrm{X}} \propto
D(R)^{-0.77}$, these changes in $D(R)$ lead to only modest increases
in the $2$--$10$~\kev\ flux. Interestingly, strong Comptonization effects when
$\tau_{\mathrm{T}} \ga 20$ will also moderately increase the soft flux even if
$h_f$ is held constant. We conclude that an AGN with a constant
density warm corona subject to a variable $h_f$
will move horizontally through this flux-flux plot, while changes in
$\tau_{\mathrm{T}}$ will cause slight changes in the vertical
direction. An example of such a path is shown as the solid black like
which is measured from the four spectra shown in
Fig.~\ref{fig:constdens}.

A similar pattern is seen in Fig.~\ref{fig:scatterconstdens}(b) which
plots $\Gamma_{\mathrm{2-10\ keV}}$ against the soft X-ray flux. The
model calculations assume a fixed $\Gamma=1.9$, and only minor changes
from this value are observed due to changes in $h_f$ and
$\tau_{\mathrm{T}}$. The one exception is the $h_f=0.2$,
$\tau_{\mathrm{T}}=8$ model, shown in the upper-left panel of
Fig.~\ref{fig:constdens}, as its spectrum is strongly influenced by
X-ray reflection features. Consequently,
we also expect that constant density warm coronae will move largely
horizontally in this plane, driven by changes in $h_f$ (solid black
line). Changes in the warm corona heating and structural properties
will not lead to significant vertical shifts in the plot.

The lower-left panel of Figure~\ref{fig:scatterconstdens} shows that
soft excesses produced by these constant density warm coronae predict
a correlation between $\Gamma_{\mathrm{0.5-2\ keV}}$ and the
$0.5$--$2$~keV flux. Basically, as the soft excess gets stronger, it
also produces a softer spectrum. As before, the horizontal movement
through the plot is driven by the increase in $h_f$. A larger
$h_f$ and $\tau_{\mathrm{T}}$ both yield a softer
$\Gamma_{\mathrm{0.5-2\ keV}}$, as each contribute to the importance
of Compton cooling in the slabs. 

Finally, panel (d) of Fig.~\ref{fig:scatterconstdens} shows an
interesting separation between low and high $\tau_{\mathrm{T}}$ models
in a plot of $\Gamma_{\mathrm{0.5-2\ keV}}$ versus the
$2$--$10$~\kev\ flux. As noted above, the hard X-ray flux is primarily
affected by $D(R)$, so low $\tau_{\mathrm{T}}$ models are on the left
hand side of the plot and more Thomson thick ones are on the right
side. Therefore, changes in accretion rate will
move an AGN horizontally through this plot. The extent of the vertical
motion due to any changes in $h_f$ depends on
$\tau_{\mathrm{T}}$. Increasing the dissipation in a thick layer
enhances Compton cooling, which leads to a softer and stronger soft
excess. These effects are weaker for lower values of
$\tau_{\mathrm{T}}$ as bremsstrahlung and line-cooling processes can
take on larger roles in the cooling of the slab
\citep[e.g.,][]{ball20}. Indeed, these effects can lead to significant
changes in the spectrum when $\tau_{\mathrm{T}}=8$ and the slab is not
dominated by Compton cooling. An AGN that has both a variable
$\tau_{\mathrm{T}}$ and $h_f$ in response to changes to $D(R)$ (e.g.,
Fig.~\ref{fig:constdens})  would
then trace out a diagonal path through this plane (black solid line).

In summary, we find that constant density models can maintain
appropriate warm coronae (with surface temperatures in the range
$0.3$--$1.1$~\kev) through changes in $D(R)$ when both the density and
optical depth of the corona increase with $D(R)$. In addition, $h_f$
will likely also increase with $D(R)$ if the soft excess is to maintained
through large increases of $D(R)$. These changes in the structural and
heating properties of warm coronae are reflected in the predicted
spectral shapes. Comparing the observed changes of AGN soft excesses
with the tracks predicted in Fig.~\ref{fig:scatterconstdens} will
provide an important test for the constant density warm corona model.

\subsection{Hydrostatic Models}
\label{sub:hydro}
While useful, constant density slabs are clearly idealizations of the
density structure at the surface of accretion discs. Therefore, it is
important to consider how a warm corona can be sustained under a more
realistic density profile. \citet{ball20} showed that the rapidly falling
density within a hydrostatic atmosphere will reduce the rate of 2-body
cooling processes, and therefore the response of a warm corona to
changes in accretion rate could be qualitatively different from the
constant density models described above.

We repeat the experiment from Sect.~\ref{sub:nhandtau} with
irradiated and heated hydrostatic atmospheres assumed to be at the
surface of a radiation-pressure dominated accretion disc \citep{sz94}. As
before, we define a baseline model with $\tau_{\mathrm{T}}=8$ and
$f_{\mathrm{X}}=0.1$, where $\tau_{\mathrm{T}} \propto D(R)^{3/5}$ and
$f_{\mathrm{X}} \propto D(R)^{-0.77}$. The density at the base of the
atmosphere is initially set by the \citet{sz94} radiation-pressure dominated
disc equations, but the density structure is ultimately determined by
requiring the atmosphere to be in hydrostatic balance \citep{brf01}. The base densities
therefore depend both on $D(R)$ and $h_f$, and typically vary over a
small range for the models with interesting warm corona with values
$\sim 10^{13}$~cm$^{-3}$. To compensate for the lower densities, the
initial dissipation flux used in the hydrostatic models is set to
$D(R)_{i,H}=3.59\times 10^{15}$~erg~cm~$^{-2}$~s$^{-1}$, equivalent to
the accretion flux produced at a radius of $5$~Schwarzschild radii
around a $3\times 10^7$~M$_{\odot}$ black hole accreting at $5$\% of
its Eddington rate. The emission
and reflection spectra are then computed for atmospheres with
$D(R)/D_{i,H}=1,2$ and $4$, varying $h_f$ between $0.1$ and $0.8$ for
each value of $D(R)$.

\begin{figure*}
  \includegraphics[width=0.98\textwidth]{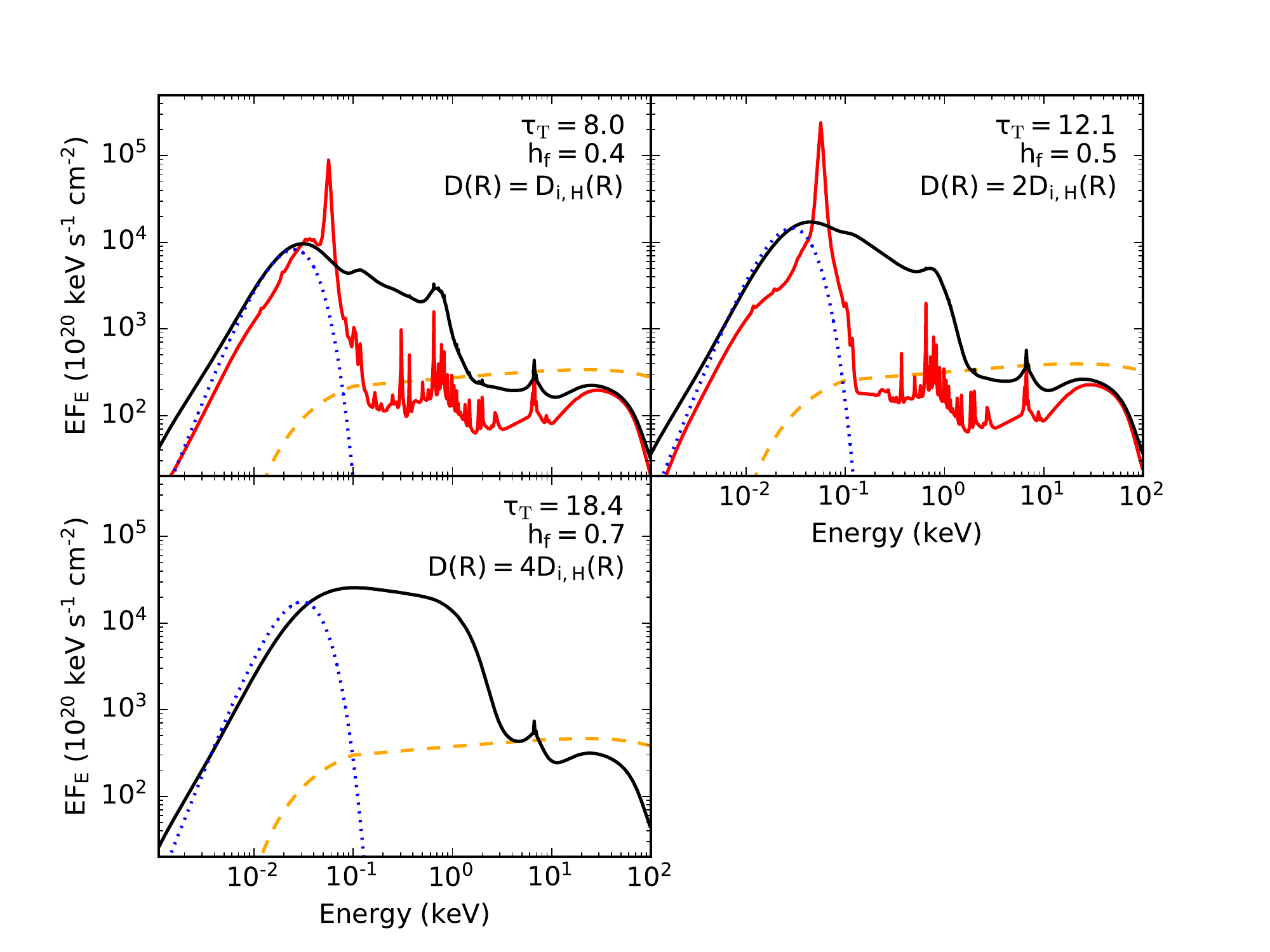}
  \caption{As in Figure~\ref{fig:constdens}, but now the modeled layer
    is a hydrostatic atmosphere at the surface of a radiation-pressure
    dominated accretion disc \citep{sz94}. The baseline accretion flux for
    this set of calculations is $D_{i,H}(R)=3.59\times
    10^{15}$~erg~cm$^{2}$~s$^{-1}$. The line styles are the same as in
    Fig.~\ref{fig:constdens}. The $\tau_{\mathrm{T}}=18.4$ calculation
    with no warm coronal heating failed because the gas pressure
    fell too low at the base of the layer; hence, there is no solid
    red line in this panel. All three models produce strong
    soft excesses with the $\tau_{\mathrm{T}}=18.4$ spectrum showing
    an extreme example of the impact of strong heating in the warm
    corona. This spectrum is affected by a hot `skin' formed at the
    surface of the atmosphere which further upscatters the
    radiation escaping from the atmosphere.  }\label{fig:hydro}
\end{figure*}

Figure~\ref{fig:hydro} shows three examples of the spectra produced by
the irradiated and
heated atmospheres, one at each value of $D(R)$, with all producing a
strong, smooth soft excess.
For hydrostatic models, the steep fall off in density with height
causes a large increase in coronal heating and gas temperature close
to the surface (e.g., Eq.~\ref{eq:hfunct}). Therefore, it is most
useful to examine the properties of the atmospheres at
$\tau_{\mathrm{T}}=1$, which occurs close to the transition in the
density profile, and is the location which largely  determines
the properties of the emitted spectrum \citep[e.g.,][]{brf01}. At this optical depth, the gas temperature is
$\approx 0.83$~keV when $D(R)/D_{i,H}(R)=1$, $\approx 0.87$~\kev\ at
$D(R)/D_{i,H}(R)=2$ and $\approx 1$~\kev\ when
$D(R)/D_{i,H}(R)=4$. All these temperatures are consistent with the
warm corona temperatures inferred by \citet{petrucci18} from their
modeling of \xmm\ spectra.

The evolution of the soft excess with $D(R)$ in these hydrostatic
models appears to follow the same trends as the ones seen in the
constant density warm corona (Fig.~\ref{fig:constdens}). The soft excess is
imprinted with spectral features when
$\tau_{\mathrm{T}}=8$ which are then smoothed out at larger optical
depths, when Compton cooling processes dominate throughout the
atmosphere. However, the spectrum produced by the
$\tau_{\mathrm{T}}=18.4$ model in Fig.~\ref{fig:hydro} shows a very
extreme soft excess that extends above $2$~\kev.
\begin{figure}
  \includegraphics[width=0.48\textwidth]{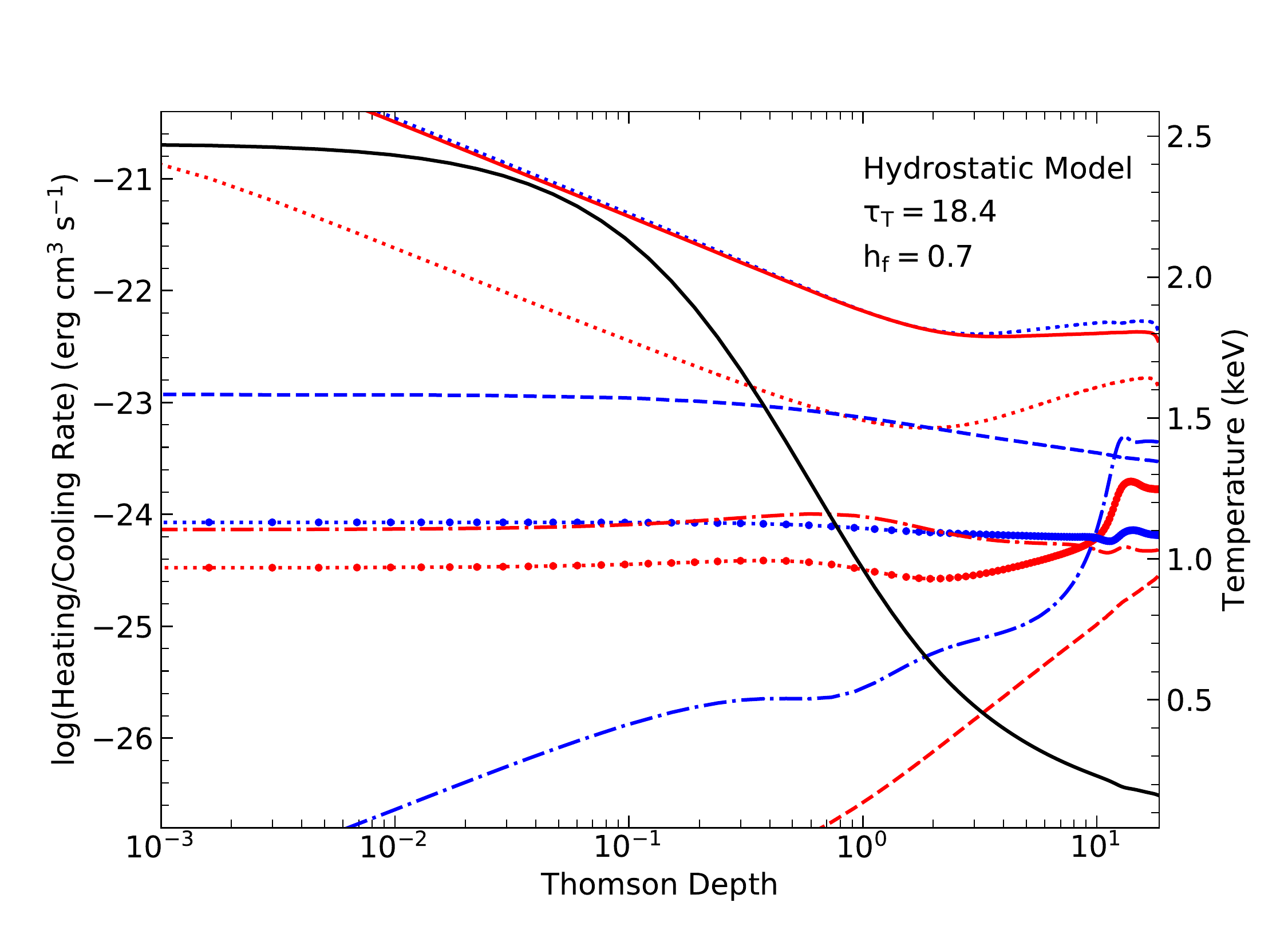}
  \caption{Details on the physical conditions of the
    $\tau_{\mathrm{T}}=18.4$, $h_f=0.7$ hydrostatic warm corona model
    whose spectrum is shown in the bottom panel of
    Fig.~\ref{fig:hydro}. The red and blue lines show the heating and
    cooling rates, respectively, as a function of Thomson depth into
    the atmosphere, and should be read using the left-hand axis.  The different processes are
    denoted by different line styles: Compton heating and cooling
    (dotted lines), bremsstrahlung heating and cooling (short-dashed
    lines), photo-ionization heating and line cooling (dot-dashed
    lines), and recombination heating and cooling (thick dotted
    lines). The coronal heating function, $\mathcal{H}$
    (Eq.~\ref{eq:hfunct}), is plotted as the solid red line. This rate
    increases steadily at $\tau_{\mathrm{T}} \la 1$ due to the drop in
    gas density. The solid black line plots the gas temperature
    through the atmosphere (read with the right-hand axis). In this
    model, a Compton dominated warm corona exists throughout the
    atmosphere, but the strong heating generates a `hot skin' at its
    surface which provides additional broadening to the emitted
    spectrum.} \label{fig:hotskin}
  \end{figure}
As seen in Figure~\ref{fig:hotskin}, this spectrum is produced by a
warm corona that is Compton dominated throughout the atmosphere, but the rapid
decline in density leads to a `hot skin' situated
at the surface \citep[e.g.,][]{nk01}. This hot skin provides an additional Compton
screen through which deeper emission must pass through. Due to
its high temperature, upscattering in the skin smears the emitted
spectrum toward higher energies. These extreme soft excesses can only occur in models where a
hot skin forms at the disc surface, and therefore are not found in
our set of constant density models\footnote{In contrast, constant
  density warm coronae will produce a Comptonized
bremsstrahlung spectrum when $h_f$ is large \citep{ball20}; however,
these models yield surface temperatures $> 1.1$~\kev\ and are not
included in our samples of soft excesses.}. As discussed
below, the hot skin spectra lead to interesting observational
properties.

\begin{figure*}
  \includegraphics[width=0.98\textwidth]{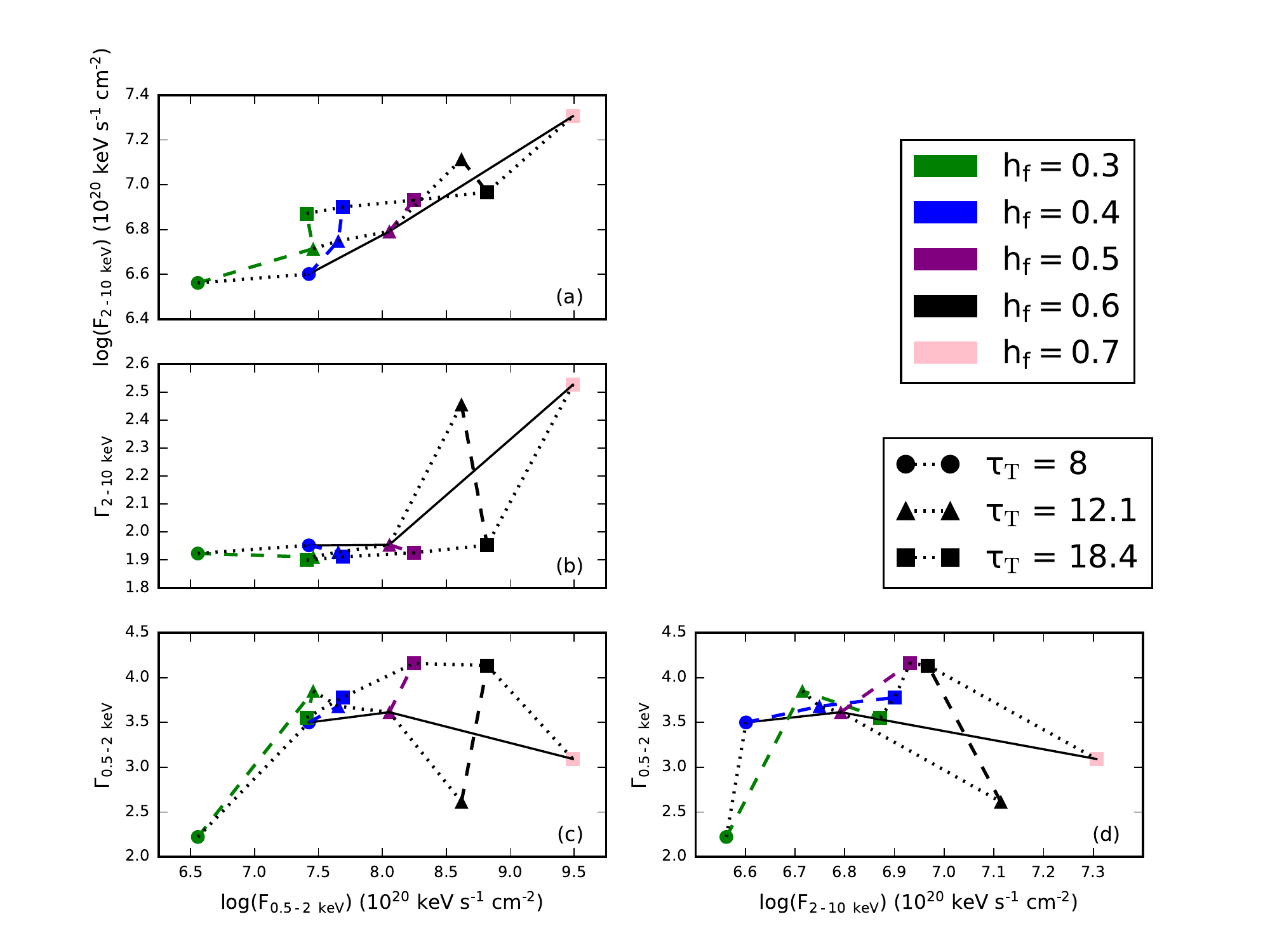}
  \caption{Trends of observational parameters in the X-ray spectra
    produced by the hydrostatic warm corona models where
    $\tau_{\mathrm{T}}$ and $f_{\mathrm{X}}$ are
    functions of $D(R)$. Only models with a
    temperature at $\tau_{\mathrm{T}}=1$ between $0.3$ and $1.1$~\kev\ are included in
    this analysis. The photon indices and fluxes are
    calculated from spectra that combine the predicted
    emitted/reflected spectrum (black solid lines in
    Fig.~\ref{fig:hydro}) with the illuminating power-law (dashed
    lines in Fig.~\ref{fig:hydro}), assuming a reflection fraction
    of $0.7$. The different symbols denote
    the results from the different $\tau_{\mathrm{T}}$ while the
    colours indicate the value of $h_f$ used in the model. Dotted lines
    in each panel connect results with the same $\tau_{\mathrm{T}}$,
    and the coloured dashed lines connect results with the same
    $h_f$. The three models shown in Fig.~\ref{fig:hydro} follow the
    solid black line in each panel. The fluxes are
    calculated in units of emitting area.  } \label{fig:scatterhydro}
\end{figure*}

Figure~\ref{fig:scatterhydro} shows the observational
characteristics of the emission and reflected spectra produced by the
hydrostatic warm corona models. The layout of the panels is the same as
the one for the constant density calculations
(Fig.~\ref{fig:scatterconstdens}). 
The hydrostatic atmospheres plotted in the Figure are selected to
all have temperatures at $\tau_{\mathrm{T}}=1$ between $0.3$ and
$1.1$~\kev\ to ensure they bracket the inferred properties of warm
absorbers \citep[e.g.][]{petrucci18}. As in
Fig.~\ref{fig:scatterconstdens}, the measurements are made from
spectra that have a reflection fraction of $0.7$.

Many of the trends observed in Fig.~\ref{fig:scatterhydro} are similar
to the ones in Fig.~\ref{fig:scatterconstdens}. For example, in panel
(a), horizontal motion across the figure is driven by
increasing $h_f$, while increases in the $2$--$10$~\kev\ flux, which
drives vertical motion, is largely achieved by increasing $D(R)$ (and,
hence, larger $\tau_{\mathrm{T}}$). However, there are two interesting
exceptions. The first is a significant increase in the soft X-ray flux
in the $\tau_{\mathrm{T}}=8$ models when $h_f$ is changed from $0.3$
to $0.4$ due to the development of a Compton cooled warm corona. The
second exception is the large jump in $F_{\mathrm{2-10\ keV}}$ due to
the $h_f=0.7$, $\tau_{\mathrm{T}}=18.4$ model (bottom spectrum in
Fig.~\ref{fig:hydro}). The extreme soft excess in that spectrum
actually extends above $2$~\kev\ enhancing the flux in the hard X-ray
band. An AGN that is subject to significant heating in the warm corona
may therefore trace a steep trajectory in this flux-flux plane (e.g.,
the solid black line, which follows the three models shown in
Fig.~\ref{fig:hydro}).

The hard X-ray photon index, $\Gamma_{\mathrm{2-10\ keV}}$, predicted
by the hydrostatic warm corona models is relatively insensitive to the
affects of the corona heating (Fig.~\ref{fig:scatterhydro}(b)), with
values close to the input slope of $1.9$. This result is consistent
with the results of the constant density models
(Fig~\ref{fig:scatterconstdens}). Interestingly, there are two
significant outliers to this trend which both show
$\Gamma_{\mathrm{2-10\ keV}} \approx 2.5$. These two warm corona
generate extreme soft excesses that extends to energies $>
2$~\kev\ due to the formation of a `hot skin' that upscatters the
emitted radiation field (e.g., the
bottom panel of Fig.~\ref{fig:hydro}). Thus, even though the irradiating
power law has a $\Gamma=1.9$, the inferred
$\Gamma_{\mathrm{2-10\ keV}}$ would be significantly steeper.

Panel (c) of Fig.~\ref{fig:scatterhydro} shows a more complex relationship between
$\Gamma_{\mathrm{0.5-2\ keV}}$ and $F_{\mathrm{0.5-2\ keV}}$ for the
hydrostatic warm coronae than shown in
Fig.~\ref{fig:scatterconstdens}(c) for the constant density models. We again
  see a correlation between the photon index and the soft X-ray flux
  that traces the development of a Compton cooled warm corona. Once
  this is established, then increases in $h_f$ only changes
  $\Gamma_{\mathrm{0.5-2\ keV}}$ by a small amount until the
  temperature of the warm corona increases to the point that
  the hot skin forms and contributes to the soft excess. Once this
  stage is achieved, the soft X-ray photon index begins to drop with
  X-ray flux, as the shape of the spectrum is flattened by
  Comptonization in the hot skin. Thus, an AGN that
  follows a declining trajectory in this plane may be showing evidence
  for the development of a strongly heated warm corona (e.g., the
  solid black line in panel (b)).

  A similar effect is seen in the final panel of
  Fig.~\ref{fig:scatterhydro} which plots
  $\Gamma_{\mathrm{0.5-2\ keV}}$ against the $2$--$10$~\kev\ flux
  calculated from the hydrostatic warm corona models. There is little
  change in $\Gamma_{\mathrm{0.5-2\ keV}}$ once a Compton cooled warm
  corona develops, but a large change in the photon index and
  $F_{\mathrm{2-10\ keV}}$ occurs when an extreme soft excess is
  produced by strong coronal heating. As described above (and seen in
  the bottom panel of Fig.~\ref{fig:hydro}), the soft excess in this
  case is impacted by Comptonization in a hot skin that
  flattens the soft X-ray spectrum and extends it into the hard X-ray
  band. Thus, these models will lead to an increase in
  $F_{\mathrm{2-10\ keV}}$ accompanied by a drop in
  $\Gamma_{\mathrm{0.5-2\ keV}}$. This change in slope occurs faster
  when coronal heating increases in atmospheres with smaller $\tau_{\mathrm{T}}$
  (comparing the triangles to the squares in the panel (d)).

  The above results show that a strong soft excess produced by a 
  hydrostatic warm corona can be sustained through large changes in
  dissipated flux by appropriate increases in the optical depth and
  $h_f$. The rapidly declining density structure of the hydrostatic
  atmosphere allows the production of more extreme soft excesses when
  the coronal heating is large. These soft excesses still have
  temperatures at $\tau_{\mathrm{T}}=1$ of $\sim 1$~\kev\ but produce a
  broader and flatter soft excess (due to Comptonization in a hot skin) that extends above $2$~\kev. The resulting spectra will be
  observed to have a steep $\Gamma_{\mathrm{2-10\ keV}}$, and their
  development will lead to negative correlations between
  $\Gamma_{\mathrm{0.5-2\ keV}}$ and the observed soft and hard fluxes.

\section{Discussion}
\label{sect:discuss}
Recent work by \citet{ball20} and \citet{petrucci20} have shown that a
warm corona with $\tau_{\mathrm{T}} \approx 10$--$20$ at the surface
of an accretion disc can, under certain conditions, produce a strong,
smooth soft excess that is qualitatively consistent with
observations. This paper explores the question of how the warm corona
and heating properties change in order to sustain the soft excess
while the underlying accretion flux, $D(R)$,  varies. The results shown above
show that maintaining a warm corona of the right properties can be
accomplished if $\tau_{\mathrm{T}}$, and, to a lesser extent,
$n_{\mathrm{H}}$ increases with $D(R)$. The fraction of
accretion energy dissipated in the warm corona, $h_f$, may also change
with $D(R)$, depending on the value of $\tau_{\mathrm{T}}$. The range
of possible $h_f$ that yields a warm corona is narrower
for smaller $\tau_{\mathrm{T}}$. Thus, it is easier to maintain a warm
corona when $\tau_{\mathrm{T}}$ is large (e.g.,
Fig.~\ref{fig:scatterconstdens}). In this section, we compare the
results shown in Sect.~\ref{sect:results} to recent observations measuring
changes in AGN soft excesses. As the warm coronae models are
calculated only at a single disc radius, the comparison to
observations will necessarily be qualitative. Model accretion disc
spectra incorporating both a hard X-ray power-law and a warm corona is
planned for future work.

Figures~\ref{fig:scatterconstdens} and~\ref{fig:scatterhydro} show how
the observed properties of soft excesses predicted by the warm corona
models vary with flux. These soft excesses are produced by the
combination of Comptonized blackbody emission passing through the warm
corona, plus bremsstrahlung and recombination line radiation due to
illumination from the external hard X-ray power-law. Both panels
assume the final spectrum has a fixed reflection fraction of
$0.7$. Figure~\ref{fig:scatterconstdens} shows the changes in the soft
excess as the warm corona properties are modified to be sustained over
a factor of 8 increase in $D(R)$. The soft excess grows stronger and
softer as the accretion flux increases, with the strength of the
dependence varying on the details of $h_f$ (e.g.,
Fig.~\ref{fig:scatterconstdens}). However, these changes often have a
very minor effect on the shape of the $2$--$10$~\kev\ spectrum.  In contrast,
Fig.~\ref{fig:scatterhydro} shows how a strongly heated warm corona
can produce an extreme soft excess that extends above $2$~\kev. In
this scenario, the stronger and softer relationships seen in the
previous case inverts, and the spectrum becomes both harder on the
soft band and softer in the hard band with a
$\Gamma_{\mathrm{2-10\ keV}} \approx 2.5$. These extreme soft excesses are
produced in strongly heated hydrostatic atmosphere due to the
formation of a hot skin at the surface which further Compton upscatters
the emitted spectrum. The hard X-ray spectra of many AGNs appear to
soften as their flux increases \citep[e.g.,][]{kb16}, and there is
tentative evidence that $\Gamma_{\mathrm{2-10\ keV}}$ is correlated
with the Eddington ratio \citep[e.g.,][]{trak17}. The formation of this hot skin and
its dependence on $h_f$ is a possible explanation for these
relationships. If more accretion energy is dissipated in a warm corona
at larger Eddington ratios than a `hot skin' will also develop at
larger accretion rates. Comptonization in this hot skin will steepen
the observed $2$--$10$~\kev\ spectrum, independent of the shape of the
illuminating power-law. The effectiveness of the process will depend
on the details of the heating in both the warm and hot coronae, and
could naturally account for the observed scatter in the relationships
between photon index and Eddington ratio. The impact of the
extreme soft excess on the spectrum will be less important at energies
$\ga 10$~\kev, and the spectrum will appear to harden at these
energies. Therefore, a power-law fit to such a spectrum (e.g., one
observed by \nustar) will return a large reflection fraction. AGNs
with steep power-laws and strong reflection fractions
\citep[e.g.,][]{jiang18} may therefore be explained by a strongly
heated hydrostatic warm corona. 

AGNs with soft excesses will trace out trajectories through the panels of
Figs.~\ref{fig:scatterconstdens} and~\ref{fig:scatterhydro}. By comparing these
trajectories to the different relationships plotted in the figures, we
can determine if these trajectories are consistent with a warm corona
model that has a constant $\tau_{\mathrm{T}}$ or $h_f$, or one that
has a variable structure. For example, \citet{ursini20} recently
studied five joint \xmm\ and \nustar\ observations of the `bare'
Seyfert 1 galaxy HE~1143-1810. All five observations of the AGN show a
strong soft excess, and the five observations span a factor of
$\approx 2$ in flux. \citet{ursini20} considered a two corona model that
includes emission from both a hot and warm Comptonizing corona. The
authors found that the photon index of the warm corona was
anti-correlated with the soft flux, and there was also a correlation
between the hard and soft fluxes. A correlation between the soft and
hard band fluxes is a common prediction of the warm corona models
described here,
with a shallow slope for a constant $\tau_{\mathrm{T}}$, and a steeper
one if $\tau_{\mathrm{T}}$ is changing (e.g., Fig.~\ref{fig:scatterhydro}(a)). Interestingly, an
anti-correlation between the soft photon index and soft flux is only
seen in Fig.~\ref{fig:scatterhydro}(c) when $\tau_{\mathrm{T}}$ is
constant and $h_f$ is increasing (e.g., the triangles in
Fig.~\ref{fig:scatterhydro}(c)). \citet{ursini20} note that there is
very little change in the inferred $\tau_{\mathrm{T}}$ of the warm
corona between the low and high flux states (with $\tau_{\mathrm{T}}
\approx 17$--$18$). Therefore, qualitatively, the soft excess of
HE~1143-1810 is consistent with a hydrostatic warm corona at a fixed
$\tau_{\mathrm{T}}$ with $h_f$ increasing with flux.

Similar results were found by \citet{middei19} who analyzed 5
\xmm\ and \nustar\ observations of the Seyfert galaxy NGC 4593 with a
two corona model. These data spanned approximately a factor of 2 in
flux, and \citet{middei19} determined that the soft X-ray photon
index was anti-correlated with the hard X-ray ($2$--$10$~\kev) flux,
but, as with HE~1143-1810, the hard and soft band fluxes were
correlated. Fig.~\ref{fig:scatterhydro}(a) and (d) indicate that
evolution at a fixed $\tau_{\mathrm{T}}$ is consistent with these
trends. Interestingly, \citet{middei19} also find an anti-correlation
between the photon indices of the soft and hard
components.
\begin{figure*}
  \includegraphics[width=0.98\textwidth]{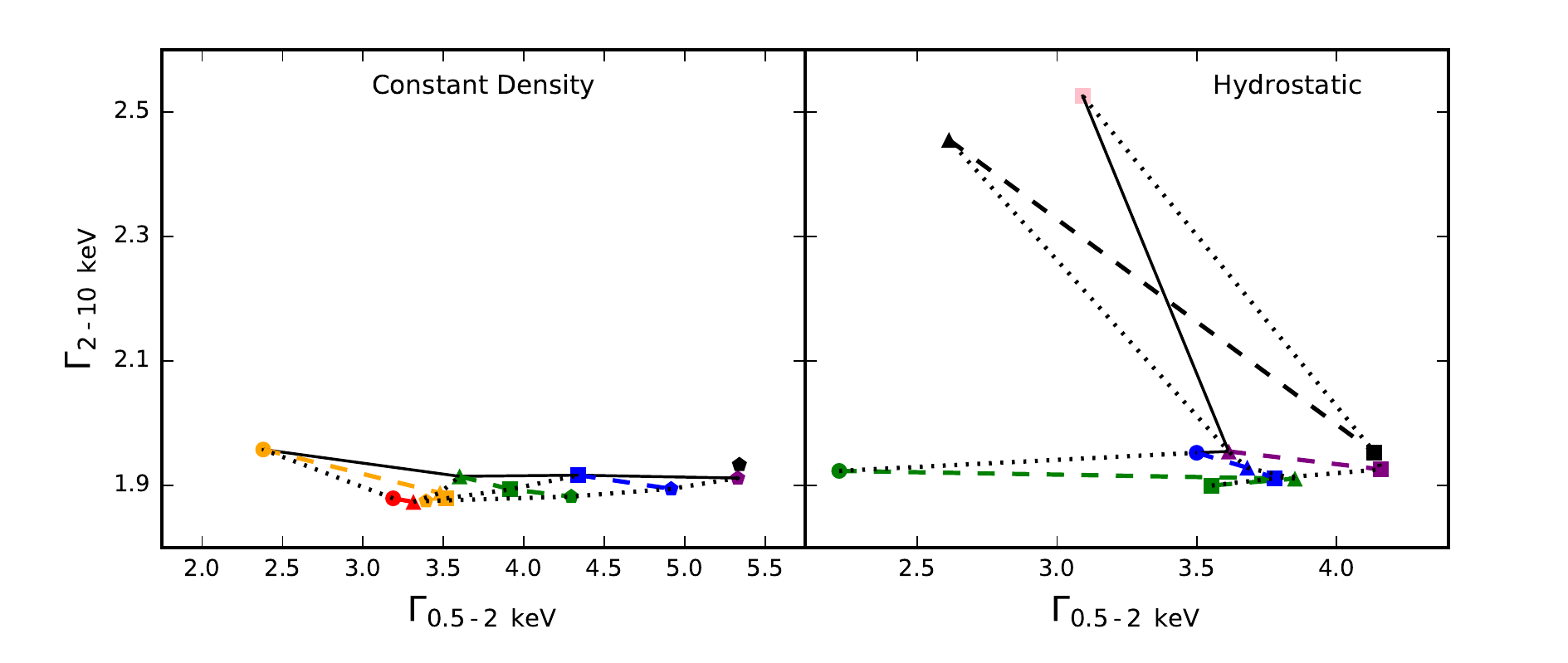}
  \caption{Potential relationships between the hard and soft X-ray
    photon indices measured from the constant density (left-hand
    panel) and hydrostatic (right-hand panel) warm corona models. The
    plotted points and lines are the same as the ones from
    Figs.~\ref{fig:scatterconstdens} and \ref{fig:scatterhydro}. The
    constant density warm corona models with temperatures between
    $0.3$ and $1.1$~\kev\ produce only very small changes in
    $\Gamma_{\mathrm{2-10\ keV}}$, even if the soft X-ray photon index
    varies
    significantly. In contrast, if a hot skin forms on the surface of
    a hydrostatic model (e.g., Fig.~\ref{fig:hotskin}), then this will
    significantly alter the observed $\Gamma_{\mathrm{2-10\ keV}}$,
    even though the spectrum emitted by the hot corona is
    unchanged.} \label{fig:gammas}
\end{figure*}
Fig.~\ref{fig:gammas} shows that no relationship between the photon
indices is found from the constant density warm corona models with
surface temperatures of $0.3$--$1.1$~\kev, despite large changes in
$\Gamma_{\mathrm{0.5-2\ keV}}$ due to an increasing soft
excess. However, an anti-correlation between the two photon-indices
can be seen in hydrostatic warm coronae with a fixed
$\tau_{\mathrm{T}}$ (right-hand panel of Fig.~\ref{fig:gammas}), due
to the development of the `hot skin' at the surface of the atmosphere
(e.g., Fig.~\ref{fig:hotskin}). Although a quantitative comparison is
not yet possible, such an anti-correlation may be an indication of a
changing coronal heating fraction. However, \citet{middei19} do not
find any correlations (neither positive nor negative) between the soft
photon index and the soft X-ray flux. Given the other observed
correlations seen from this AGN, it is challenging to explain this
lack of correlation with the simple warm corona
models considered here. Clearly, there is a need for models that encompasses a larger
fraction of the accretion disc with which to compare against
observations.

\begin{figure*}
  \includegraphics[width=0.98\textwidth]{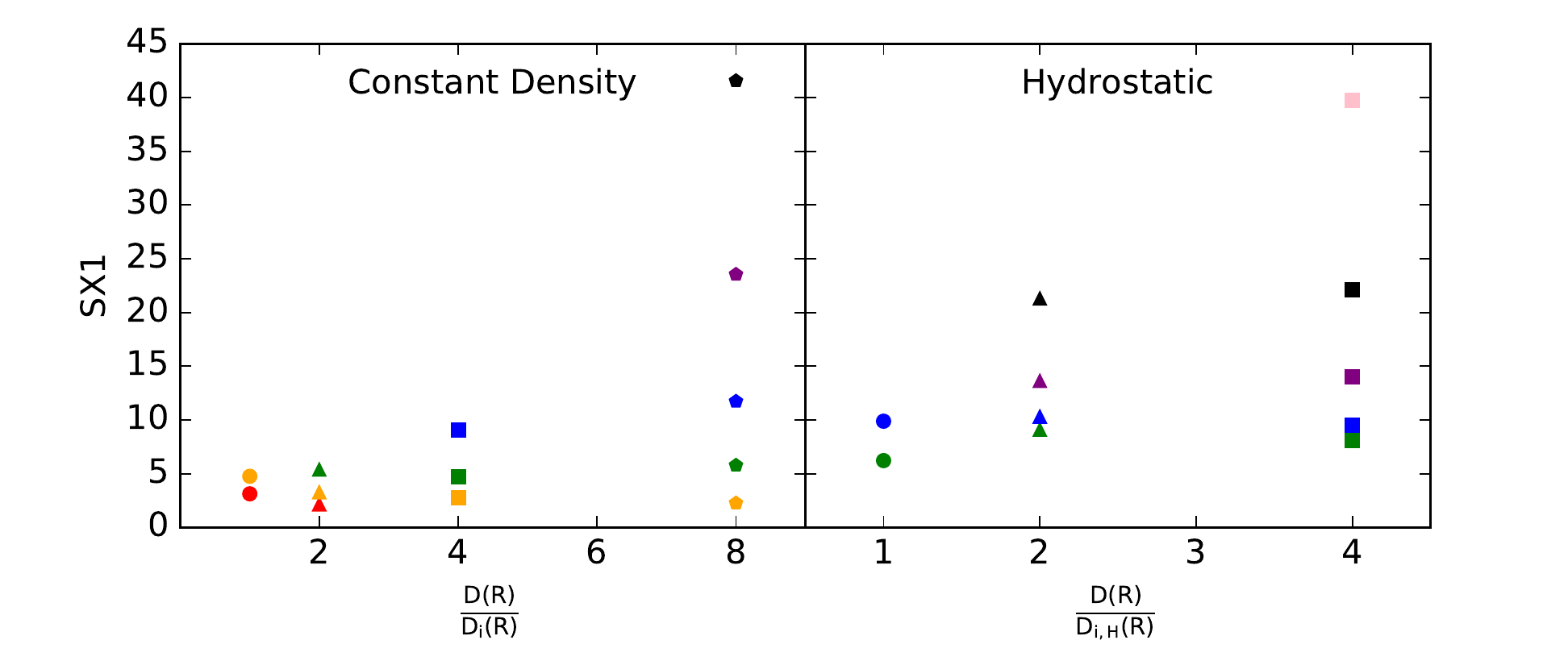}
  \caption{(Left) Plot of SX1 (Eq.~\ref{eq:sx1}) versus
    $D(R)/D_{i}(R)$ for the constant density warm corona models
    shown in Fig.~\ref{fig:scatterconstdens}. The colours and shapes
    of the points are the same as in that Figure. (Right) As in the
    other panel, but now for the hydrostatic warm corona models shown
    in Fig.~\ref{fig:scatterhydro}. In both cases, the value of SX1
    can vary significantly at a given accretion flux, depending on the
  value of $h_f$. This is consistent with observations showing a loose
  correlation between the strength of an AGN soft excess and its
  Eddington ration \citep[e.g.,][]{gw20}. } \label{fig:sx1}
\end{figure*}

A common observational measurement of the strength of a soft excess in
the spectrum of an AGN is SX1, which is defined as the
$0.5$--$2$~\kev\ flux of a blackbody fit to the soft excess, divided
by the flux of the hard X-ray power-law extrapolated into the same
band. This parameter therefore gives an estimate of the amount of
`excess' flux above what is expected from the hard X-ray power-law. A
recent analysis by \citet{gw20} of 89 type-1 AGNs find that SX1 tends
to be larger for more rapidly accreting AGNs, such as narrow-line
Seyfert 1 galaxies, albeit with a large amount of scatter. To see if
the soft excesses produced by the warm corona models also show a
similar trend we compute
\begin{equation}
  \mathrm{SX1}=\left({F^{\mathrm{refl}}_{\mathrm{0.5-2\ keV}} \over
    F^{\mathrm{pl}}_{\mathrm{0.5-2\ keV}}} \right ),
  \label{eq:sx1}
\end{equation}
where $F^{\mathrm{refl}}_{\mathrm{0.5-2\ keV}}$ is the
$0.5$--$2$~\kev\ flux of the emission/reflection spectrum emitted by
the warm corona (i.e., the black lines in Figs.~\ref{fig:constdens} and
\ref{fig:hydro}), and $F^{\mathrm{pl}}_{\mathrm{0.5-2\ keV}}$ is the
$0.5$--$2$~\kev\ flux of the power-law (i.e., the orange dashed lines
in the same figures). We compute SX1 for all the constant density and
hydrostatic warm corona models plotted in
Figs.~\ref{fig:scatterconstdens} and~\ref{fig:scatterhydro}. In
agreement with \citet{gw20}, Figure~\ref{fig:sx1} shows that SX1 is correlated with $D(R)$, the dissipated
accretion energy which is proportional to the AGN accretion rate.
Furthermore, warm corona models naturally explain the large observed
scatter in the relationship as the coronal heating fraction $h_f$ can
vary significantly from source to source. Because a broader range of
$h_f$ is allowable in warm coronae with larger $D(R)$ (i.e., a larger
$\tau_{\mathrm{T}}$), then the scatter of SX1 values is larger at
higher accretion rates, than at lower ones, in agreement with the
results of \citet{gw20}. Therefore, warm coronae with $\tau_{\mathrm{T}}$
correlated with the accretion rate (and a variable $h_f$), appear to
be a viable explanation for the observed trends in the soft excess strength.

\section{Conclusions}
\label{sect:concl}
Any model explaining the origin of the soft excess in AGNs must be
able to account for its ubiquity, its persistence to changes in
luminosity, and both its short and long time-scale variability. A soft
excess produced by a combination of a warm corona and reflected
emission has many qualities consistent with these
properties. Reflection from an external hard X-ray power-law will
naturally produce a soft excess that can then be strengthened and
smoothed by Comptonized thermal emission from a warm corona. Thus, as
long as the hard X-ray and optically thick accretion disc are present,
a soft excess should be a common feature in AGNs. As shown above, variations in the
external X-ray flux, and, in particular, in the structure and heating
of the warm corona, will sustain the soft excess through changes in
the underlying accretion rate, and drive variations in the observed
spectra (e.g., Figs.~\ref{fig:scatterconstdens}
and~\ref{fig:scatterhydro}). AGN observing campaigns that trace the changes
in the soft excess spectral properties with flux will be able to
constrain the heating properties of the warm corona. For example,
comparing the warm corona models with the soft excess changes seen in HE~1143-1810
\citep{ursini20} and NGC~4593 \citep{middei19} indicates that a
variable heating fraction in a hydrostatic warm corona is the primary driver of 
the observed changes. More rigorous results will be obtained by using
datasets that span a larger range of flux variations, perhaps by
exploiting archival data.

Significant work remains in order to determine if warm coronae are an
important contributor to the soft excess of all AGNs, or are only
relevant for a smaller population of sources (e.g., those that exhibit
ionized \fe\ lines). The one-dimensional
models described here and by \citet{petrucci20} need to be extended
to account for emission from a larger fraction of the accretion
disc before a rigorous comparison to AGN spectra can be
performed. Similarly, additional theoretical and numerical work on the
structure and heating of the outermost layers of accretion flows is needed to inform the construction of spectral
models such as the ones presented in this paper. For example, magnetic heating
\citep[e.g.,][]{gr20} will provide a different vertical heating
profile than the uniform heating function used here
(Eq.~\ref{eq:hfunct}), and could
affect how changes in the warm corona impact the emitted
spectrum.

Our results show that the combination of a warm corona and hard X-ray
reflection produce a diversity of soft excess shapes and
strengths. Changes to the density, optical depth and/or coronal heating
strength will sustain a soft excess through large changes in accretion
rate. Overall, our results reinforce the
idea that not only can a warm corona be an important contributor to
the AGN soft excess, but also illustrates how observations of the soft
excess can provide new insights into the physics of accretions discs
around black holes.

\section*{Data Availability}
The data underlying this article will be shared on reasonable request
to the corresponding author.





\bibliographystyle{mnras}
\bibliography{refs} 




\bsp	
\label{lastpage}
\end{document}